\documentclass[twocolumn,twoside,preprintnumbers,supercriptaddress,amsmath,amssymb,nofootinbib]{revtex4}
\usepackage{epsfig}
\usepackage{graphicx}
\usepackage{dcolumn}
\usepackage{bm}
\usepackage{fancyhdr}
\usepackage{pslatex}
\begin{document}

\title{\Large Dark Energy and Some Alternatives: a Brief Overview}

\author{J. S. Alcaniz}
\affiliation{Departamento de Astronomia, Observat\'orio Nacional, 
20921-400, Rio de Janeiro -- RJ, Brasil}

\received{\today} 

\begin{abstract}

The high-quality cosmological data, which became available in the last decade, have thrusted upon us a rather preposterous composition for the universe which poses one of the greatest challenges theoretical physics has ever faced: the so-called dark energy. By focusing our attention on specific examples of dark energy scenarios, we discuss three different candidates for this dark component, namely, a decaying vacuum energy or time-varying cosmological constant [$\Lambda(t)$], a rolling homogeneous quintessence field ($\Phi$), and modifications in gravity due to extra spatial dimensions. As discussed, all these candidates [along with the vacuum energy or cosmological constant ($\Lambda$)] seem somewhat to be able to explain the current observational results, which hampers any definitive conclusion on the actual nature of the dark energy. 
\end{abstract}

\maketitle
\thispagestyle{fancy}
\section{Introduction}

According to Einstein's general theory of relativity, the dynamic properties of a given space-time are fully determined by its total energy content. In the cosmological context, for instance, this amounts to saying that in order to understand the space-time structure of the Universe one needs to identify the relevant sources of energy as well as their contributions to the total energy momentum tensor $T_{\mu \nu}$. Matter fields (e.g., baryonic matter and radiation), are obvious sources of energy. Nevertheless, according to current observations, two other components, the so-called dark matter and dark energy (or \emph{dark pressure}), whose origin and nature are completely unknown thus far, are governing the late time dynamic properties of our Universe. In the standard lore, the later actor plays a special role because it is supposed to be driving the current cosmic acceleration.

Dark energy has been inferred from a number of independent astronomical observations, which includes distance measurements of type Ia supernovae (SNe Ia) \cite{perl, rnew, Legacy2005}, estimates of the age of the Universe \cite{age}, measurements of the Cosmic Microwave Background (CMB) anisotropies \cite{boom,wmap}, and clustering estimates \cite{calb}. While the combination of the two latter results suggests the existence of a smooth component of energy that contributes with $\simeq 3/4$ of the critical density, SNe Ia observations require such a component to have a negative pressure,  which generates repulsive gravity and accelerates the cosmic expansion. In addition to these recent observational results, the astonishing success of the inflationary Cold Dark Matter (CDM) paradigm in explaining high precision measurements of CMB anisotropy, galaxy clustering, the Ly$\alpha$ forest, gravitational lensing and other astrophysical phenomena can also be thought of as an indirect but important evidence for the existence of a dominant repulsive dark energy component  

In this short contribution we review some aspects of the so-called dark energy problem. This paper does not aim to be (and is far from being) an exhaustive or complete account of the problem (for that end we refer the reader to Ref. \cite{revde}). Here, we focus our attention on three  possible mechanisms capable of explaining the current cosmic acceleration, namely, a decaying $\Lambda$ term, the potential energy density associated with a dynamical scalar field and modifications in gravity due to extra dimensions effects. As discussed below, all these possibilities (together with the standard cosmological constant $\Lambda$) seem somewhat to explain the current observational results, which makes the nature of dark energy a completely open question nowadays in Cosmology.

\section{Cosmological Constant and Dark Energy}

There is no doubt that the simplest and most theoretically appealing candidate for dark energy is the cosmological constant ($\Lambda$), whose presence modifies the Einstein field equations (EFE) to (throughout this paper we work in units where the speed of light $c = 1$)
\begin{equation} \label{efe}
R_{\mu \nu} - \frac{1}{2}g_{\mu \nu} R = {8\pi G}T_{\mu \nu} + \Lambda g_{\mu \nu} \; .
\end{equation}
Physically, $\Lambda$ acts as an isotropic and
homogeneous source  with a constant equation of state (EoS) $w_{\Lambda} \equiv p_{\Lambda}/\rho_{\Lambda} = -1$,  where $\rho_{\Lambda} \equiv \Lambda/8\pi G$\footnote{The simplest way to see how the cosmological term may lead to an accelerating expansion in a Friedmann-Robertson-Walker (FRW) universe with pressureless matter component $\rho_m$ and $\Lambda$ is by means of the Raychaudhury equation, i.e.,  $\frac{\ddot{a}}{a} = -\frac{4\pi G}{3}\rho_m + \frac{\Lambda}{3}$ or, equivalently, $f = -\frac{GM}{r^2} + \frac{\Lambda}{3} r$. Clearly, the cosmological constant gives rise to a kind of repulsive force proportional to distance, which in principle could be the mechanism behind the cosmic acceleration evidenced by current SNe Ia observations.}. From a more fundamental viewpoint, a physical basis for the cosmological constant remained unclear until 1967, when Zel'dovich \cite{zeldo} showed that $\Lambda$ is related to the zero-point vacuum fluctuations of fields and that those must respect Lorentz invariance so that $T_{\mu \nu} = - \rho_{\Lambda}g_{\mu \nu}$. Formally, the zero-point energy of a quantum field (thought of as a collection of an infinite number of harmonic oscillators in momentum space) must be infinity. If, however, we sum over the zero-point mode energies up to a certain ultraviolet momentum cutoff $k_c$ (so that the theory under consideration is still valid), we find 
\begin{equation}
\rho_{\Lambda} \sim \hbar k_c^4 \; .
\end{equation}
By considering the above expression, we show in Table I the expected contribution to the vacuum energy for some \emph{fundamental} energy scales in nature. The ratio of the values appearing in the third column of Table I to the current observational estimate of  $\rho_{\Lambda}$, i.e., $|\rho_{\Lambda}^{obs}| \sim 10^{-10}$ $\rm{erg/cm^3}$, ranges from 46-120 orders of magnitude and is the origin of the famous discrepancy between theoretical and observational estimates for $\Lambda$, the so-called cosmological constant problem (see \cite{lambda} for a review on this topic). From these arguments one may conclude that although cosmological models with a relic $\Lambda$ term ($\Lambda$CDM) are extremely successful from the observational viewpoint, the fine-tunning problem involving the vacuum energy density and $\Lambda$ seem to hamper any definitive conclusion on the cosmological constant as the actual nature of the mechanism behind cosmic acceleration (see \cite{paddy} for a broader discussion on this issue).

\subsection{Time-varying Cosmological Constant}

A phenomenological attempt at alleviating the above problem is allowing $\Lambda$ to vary\footnote{Strictly speaking, in the context of classical general relativity any additional $\Lambda$-type term that varies in space or time should be thought of as a new \emph{time-varying field} and not as a cosmological constant.  Here, however,  we adopt the usual nomenclature of time-varying or dynamical $\Lambda$ models.}. Cosmological scenarios with a time-varying or a dynamical $\Lambda$  term were independently proposed almost twenty years ago in Ref. \cite{ozer} (see also \cite{bron, lambdat, overduin}).

In order to build up a $\Lambda$(t)CDM scenario, we first note that according to the Bianchi identities, the Einstein equations (\ref{efe}) implies that $\Lambda$ is necessarily a constant either if $T^{\mu \nu} = 0$ or if $T^{\mu \nu}$ is separately conserved, i.e., $u_{\mu}T^{\mu \nu};_{\nu} = 0$. In other words, this amounts to saying that 
\begin{enumerate}
\item vacuum decay is possible only from a previous existence of some sort of non-vanishing matter and/or radiation; 

\item the presence of a time-varying cosmological term results in a coupling between $T^{\mu \nu}$ and $\Lambda$ of the type
\begin{equation} \label{coupling}
u_{\mu}{{T}}^{\mu \nu};_{\nu} =-u_{\mu}(\frac{\Lambda g^{\mu \nu}}{8\pi G});_{\nu} \; .
\end{equation}

\end{enumerate}

\begin{center}
\begin{table}[t]
\caption{Expected contribution to the vacuum energy.}
\begin{ruledtabular}
\begin{tabular}{lcl}
Energy scale& \quad \quad \quad \quad  &\quad \quad \quad \quad \quad \quad \quad  $\rho_{\Lambda}$\\ 
\hline  \\ 
QCD & \quad \quad \quad \quad  0.3 \rm{GeV} & \quad \quad \quad \quad  $\simeq 10^{36}$ $\rm{erg/cm^3}$ \\ 
Eletroweak & \quad \quad \quad \quad  $10^2$ \rm{GeV} & \quad \quad \quad \quad  $\simeq 10^{47}$ $\rm{erg/cm^3}$ \\ 
GUT & \quad \quad \quad \quad  $10^{16}$ \rm{GeV} & \quad \quad \quad \quad  $\simeq 10^{102}$ $\rm{erg/cm^3}$ \\ 
Planck & \quad \quad \quad \quad  $10^{18}$ \rm{GeV} & \quad \quad \quad \quad  $\simeq 10^{110}$ $\rm{erg/cm^3}$ \\ 
\hline   \\
\end{tabular}
\end{ruledtabular}
\end{table}
\end{center}

To proceed further, one must now specify the function $\Lambda$(t). However, in the absence of a natural guidance from fundamental physics on a possible time variation of the cosmological constant most of the decay laws discussed in the literature are  phenomenological (based on dimensional, black hole thermodynamics arguments, among others. See, e.g., \cite{overduin} and references therein). In this regard, a still phenomenological but very interesting step toward a more realistic decay law was recently discussed in Ref. \cite{wm} (see also \cite{alcaniz}), in which the time variation of $\Lambda$ is deduced from the effect it has on the CDM evolution. The qualitative argument is the following: since vacuum is decaying into CDM particles, CDM will dilute more slowly compared to its standard evolution, $\rho_m \propto a^{-3}$. Thus, if the deviation from the standard evolution is characterized by a constant $\epsilon$, i.e.,
\begin{equation}
\label{energyCDM} \rho_m=\rho_{mo} a^{-3 + \epsilon},
\end{equation}
Eq. (\ref{coupling}) yields
\begin{equation}\label{decayv}
\rho_{\Lambda} =  \tilde\rho_{\Lambda o} + \frac{\epsilon \rho_{mo}}{3 -
\epsilon}a^{-3 + \epsilon},
\end{equation}
where $\rho_{mo}$ is the current CDM energy density, $a$ is the cosmological scale factor, $\tilde\rho_{\Lambda o}$ is an integration constant, and thermodynamic considerations restrict $\epsilon$ to be $ \geq 0$ \cite{alcaniz} (a similar decay law is also obtained from renormalization group running $\Lambda$ arguments \cite{shapiro}). Note that, differently from previous vacuum decay scenarios, in which the universe is either always accelerating or always decelerating from the onset of matter domination to today, the presence of the residual term $\tilde\rho_{\Lambda o}$ in Eq. (\ref{decayv}) makes possible a transition from an early decelerated to a current accelerating phase, as evidenced by SNe Ia observations \cite{ms} (see Fig. 1a). Note also that, in the case of vacuum decay into photons\footnote{If vacuum decays into relativistic particles, all the 3's in Eqs. (\ref{energyCDM}) and (\ref{decayv}) become 4's.}, the primordial nucleosynthesis arguments discussed in Refs. \cite{sarkar} (more specifically the bounds from the primordial mass fraction of helium $^4\rm{He}$ and the primordial abundance by number of deuterium $\rm{D/H}$) may be no longer valid since even for small values of the parameter $\epsilon$ the residual term $\tilde\rho_{\Lambda o}$ may account not only for the current cosmic acceleration but also for the recent estimates of the total age of the Universe.

\begin{figure*}[t]
\centerline{\psfig{figure=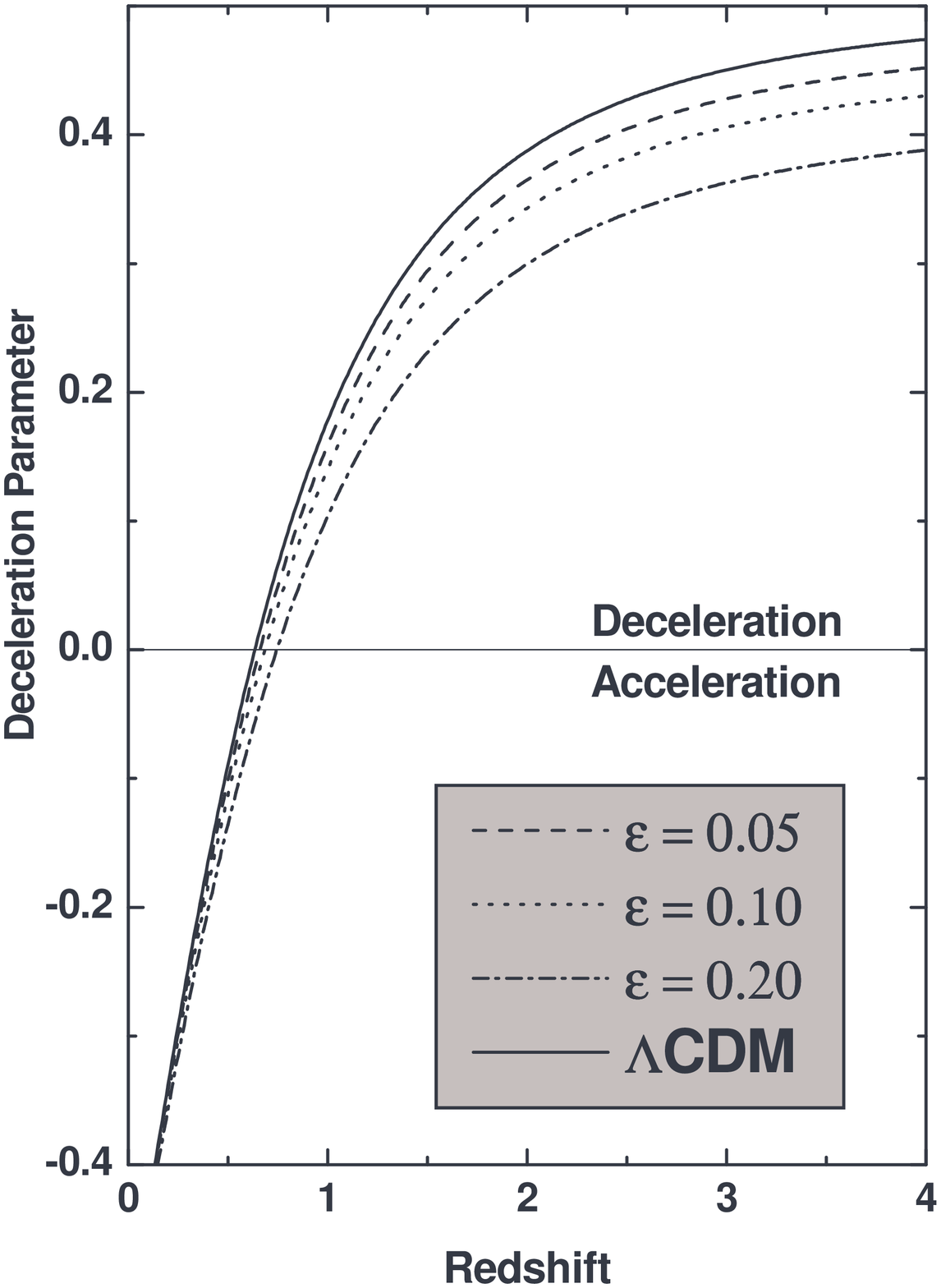,width=3.2truein,height=2.8truein,angle=0}
\psfig{figure=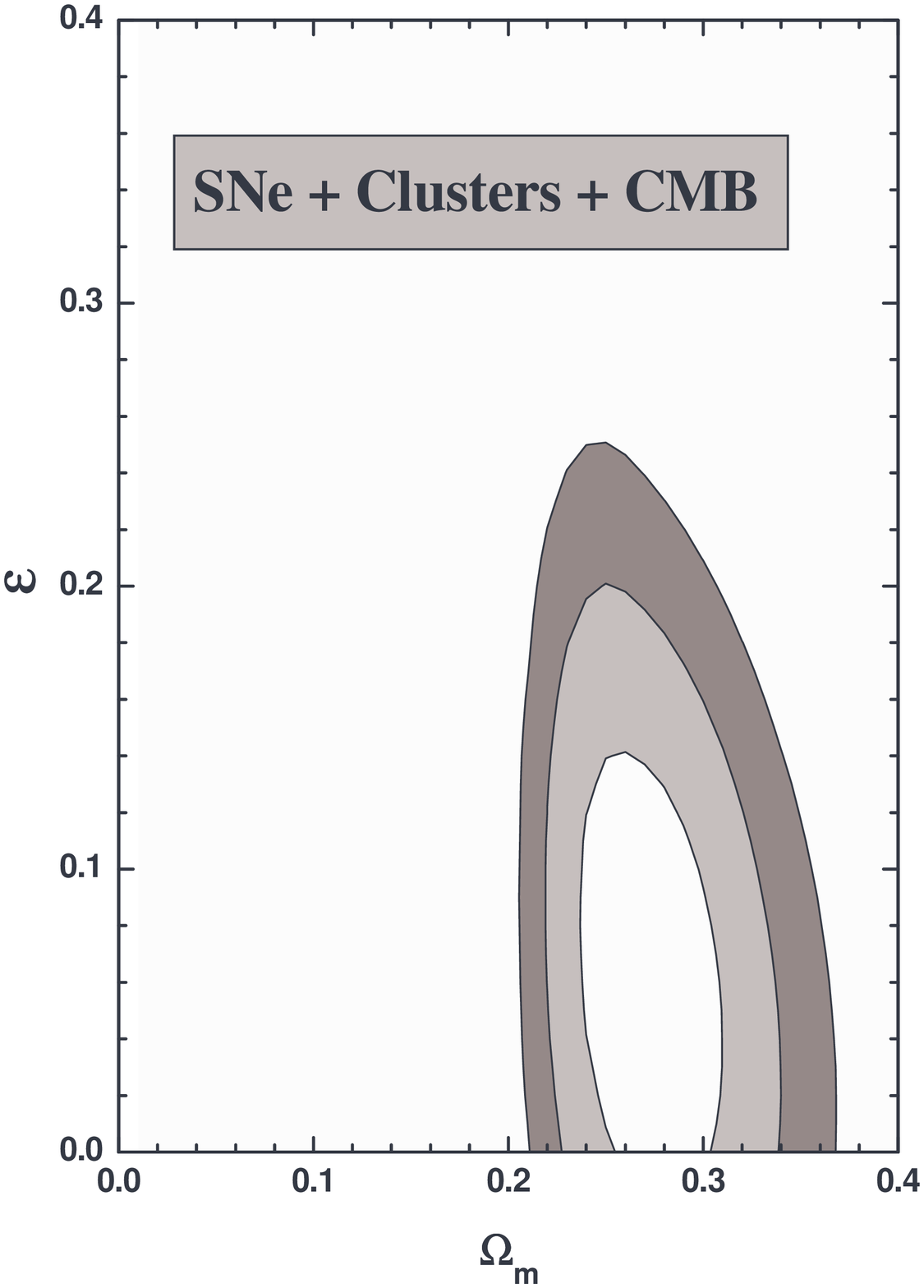,width=3.2truein,height=2.8truein,angle=0}
\hskip 0.1in} 
\caption{{\bf{Left:}}The deceleration parameter as a function of redshift for
some selected values of $\epsilon$. In all curves a baryonic
content corresponding to $\simeq 4.4\%$ of the critical density
has been considered. {\bf{Right:}} The plane $\Omega_m - \epsilon$ for the
$\Lambda(t)$CDM scenario. The curves correspond to confidence
regions of $68.3\%$, 95.4$\%$ and 99.7$\%$ for a joint analysis
involving SNe Ia, Clusters and CMB data. The best-fit parameters
for this analysis are $\Omega_m = 0.27$ and $\epsilon = 0.06$,
with reduced $\chi^2_{min}/\nu \simeq 1.14$ \emph{(Figures taken from \cite{alcaniz})}.}
\end{figure*}

If vacuum is really transferring energy to the CDM component, the immediate question one may ask is where exactly the vacuum energy is going to or, in other words, where the CDM particles are storing the energy received from the vacuum decay process. In principle, since the energy density of the CDM component is $\rho = nm$, there are at least two possibilities, namely, the current of CDM particles has a source term $N^{\alpha};_{\alpha} = \psi$ (while the proper mass of CDM particles remains constant) or  the mass of the CDM particles is itself a time-dependent quantity   $m(t) = m_oa(t)^{\epsilon}$ (while the total number of CDM particles, $N = na^{3}$, remains constant). The former possibility is the traditional approach for the vacuum decay process whereas the latter constitutes a new example of the so-called VAMP\footnote{VAriable Mass Particles}-type scenarios, in which the interaction of CDM particles with the dark energy field imply directly in an increasing of the mass of CDM particles (see, e.g., \cite{vamp} and references therein for more about VAMP models).

In Fig. 1b we compare the above $\Lambda(t)$CDM scenario with the standard one ($\Lambda$CDM), which is formally recovered for $\epsilon = 0$. We use to this end some of the most recent cosmological observations, i.e., the latest Chandra measurements of the X-ray gas mass fraction in 26 galaxy clusters, as provided by Allen et al. \cite{allen} along with the so-called \emph{gold} set of 157 SNe Ia, recently published by Riess et al. \cite{rnew}, and the estimate of the CMB shift parameter \cite{wmap}. In our analysis, we also include the current determinations of the baryon density parameter, as given by the WMAP team \cite{wmap}, i.e., $\Omega_bh^2 = 0.0224 \pm 0.0009$ and the latest measurements of the Hubble parameter, $h = 0.72 \pm 0.08$, as provided by the HST key project \cite{hst} (we refer the reader to \cite{refer} for more details on this statistical analysis). The contours stand for confidence regions ($68.3\%$, 95.4$\%$ and 99.7$\%$) in the plane $\Omega_m - \epsilon$. Note that, although the limits on the parameter $\epsilon$ are very restrictive, the analysis clearly shows that the decaying vacuum scenario discussed above constitutes a small but significant deviation from the standard $\Lambda$CDM dynamics (as mentioned earlier the standard $\Lambda$CDM model is very successful from the observational viewpoint so that large deviation from its dynamics are not expected). The best-fit parameters for this analysis are $\Omega_m = 0.27$ and $\epsilon = 0.06$, with the relative $\chi^2_{min}/\nu \simeq 1.14$ ($\nu$ is defined as degrees of freedom). Note that this value of $\chi^2_{min}/\nu$ is similar to the one found for the so-called ``concordance model" by using SNe Ia data only, i.e., $\chi^2_{min}/\nu \simeq 1.13$ \cite{rnew}. At 95.4$\%$ c.l. we also found $\Omega_m = 0.27 \pm 0.05$ and $\epsilon = 0.06 \pm 0.10$. We expect that upcoming observational data along with new theoretical developments will be able to confirm or not $\Lambda(t)$CDM models as realistic dark energy scenarios. 

\begin{figure*}[t]
\centerline{
\psfig{figure=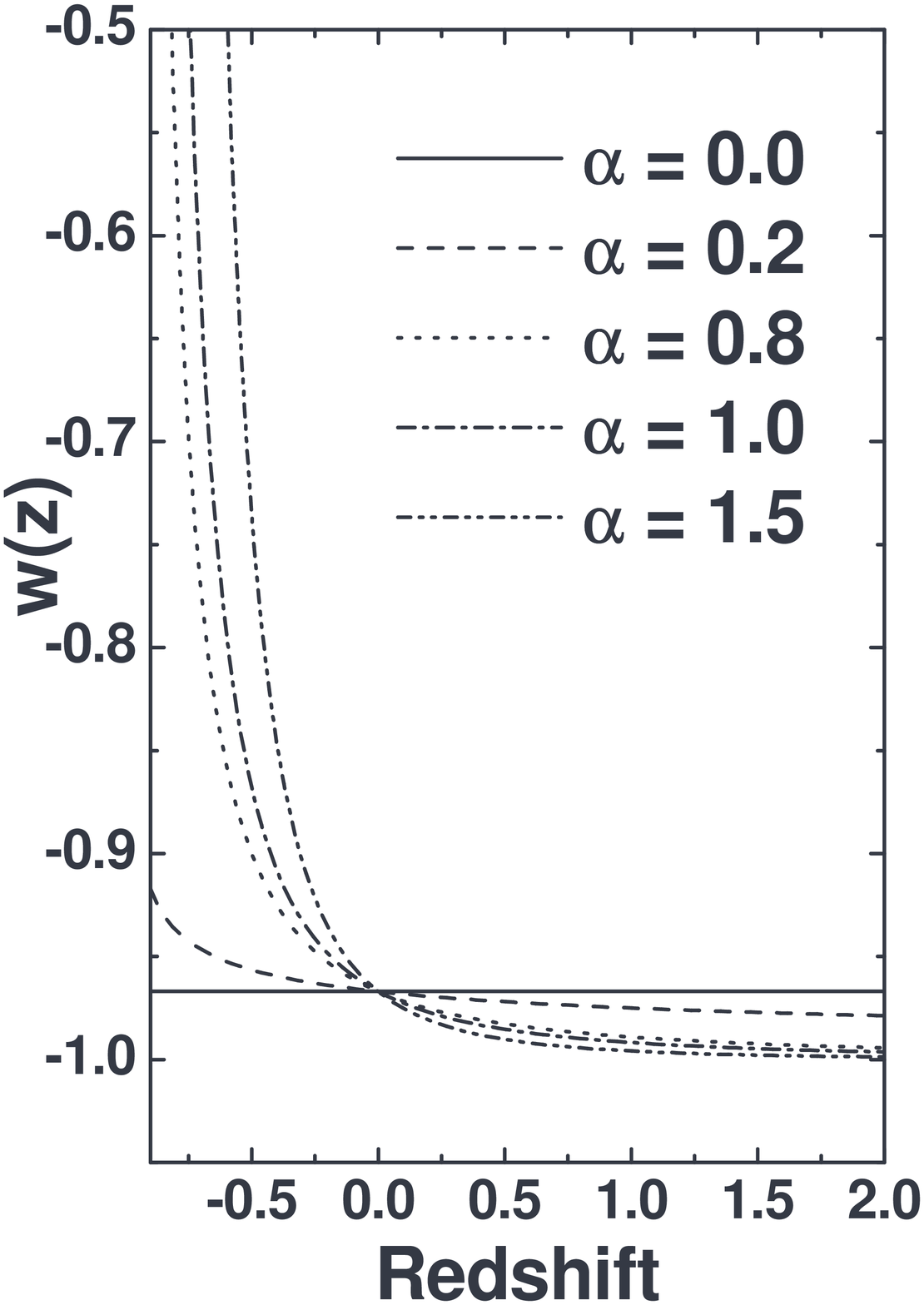,width=3.2truein,height=2.8truein}
\psfig{figure=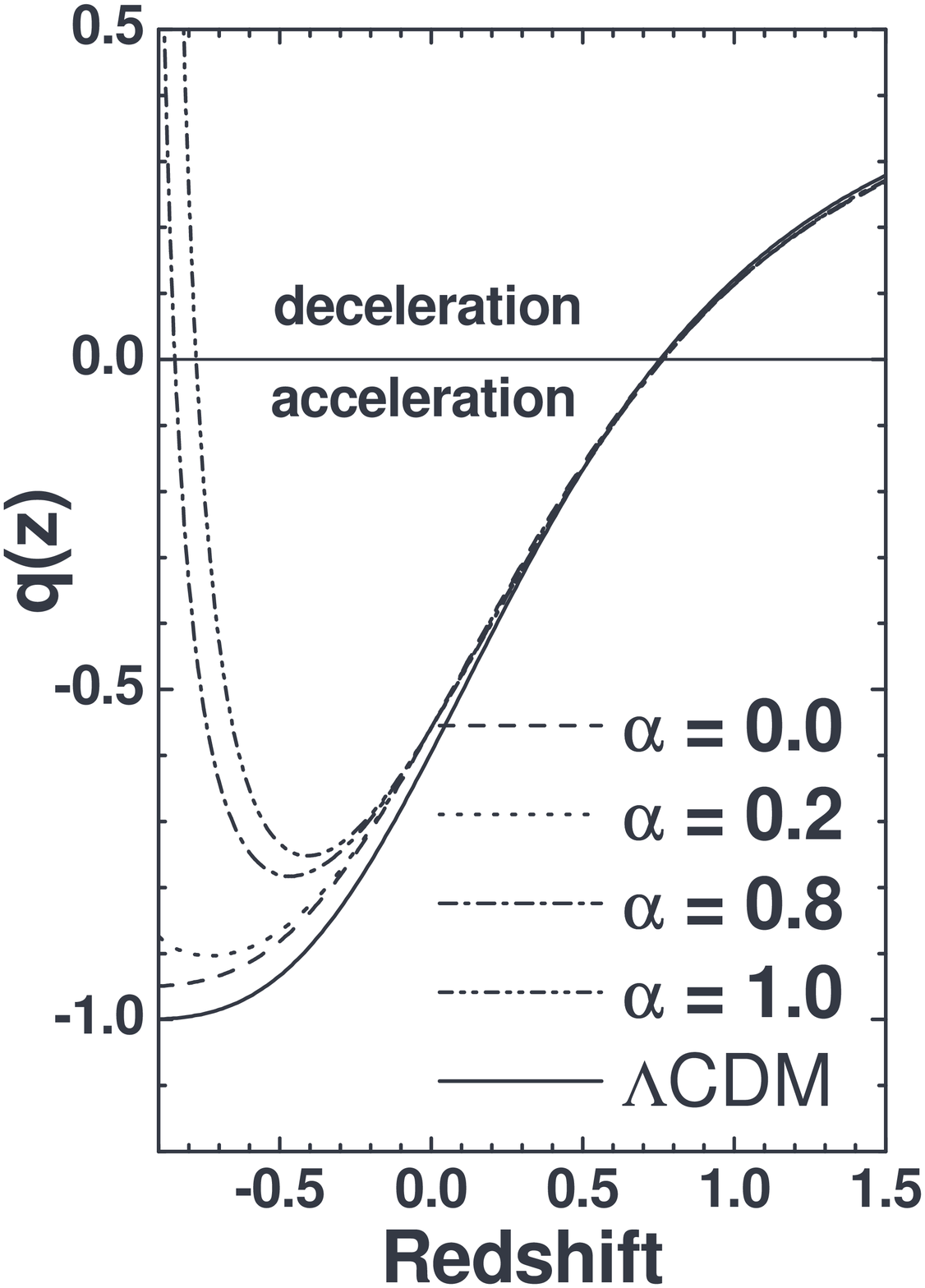,width=3.2truein,height=2.8truein} 
\hskip 0.1in} 
\caption{{\bf{Left:}} The plane $w(z) - z$. Note that $w(z)$ reduces to a constant EoS $w \simeq -0.96$ [$\lambda = {\cal{O}}(10^{-1})$] in the limit $\alpha \rightarrow 0$ while $\forall$ $\alpha \neq 0$ it was $-1$ in the past and $\rightarrow +1$ in the future. {\bf{Right:}} The deceleration parameter as a function of the redshift for selected values of $\alpha$ and $\Omega_{m,0} = 0.27$. For values of $\alpha \neq 0$ the cosmic acceleration is a transient phenomenon. In particular, for $\alpha
= 1.0$ the transition redshifts happen at $z_{a/d} \simeq \pm 0.77$ \emph{(Figures taken from \cite{prl})}. }
\end{figure*}

\section{Quintessence}

If the cosmological term is null or it is not decaying in the course of the expansion,  something else must be causing the Universe to speed up. The next simplest approach toward constructing a model for an accelerating universe is to work with the idea that the unknown, unclumped dark energy component is due exclusively to a minimally coupled scalar field $\Phi$ (quintessence field) which has not yet reached its ground state and whose current dynamics is basically determined by its potential energy $V(\Phi)$. This idea has received much attention over the past few years and a considerable effort  has been made in understanding the role of quintessence fields on the dynamics of the Universe \cite{quintessence}. Examples of quintessence potentials are ordinary exponential functions $V(\Phi) =
V_0\exp{(-\lambda \Phi)}$ \cite{RatraPeebles,fj,wetterich}, simple power-laws of the type $V(\Phi) = V_0\Phi^{-n}$ \cite{power_law},  combinations of exponential and sine-type functions $V(\Phi) = V_0\exp{(-\lambda \Phi)}[1 + A\sin{(-\nu \Phi)}]$ \cite{dodelson}, among others (see, e.g., \cite{revde,quintessence} and references therein).

An important aspect worth to emphasize is that all these quintessence scenarios are based on the premise that fundamental physics provides motivation for light scalar fields in nature so that a quintessence field $\Phi$ may not only be identified as the dark component dominating the current cosmic evolution but also as a bridge between an underlying theory and the observable structure of the Universe. If, however, it is desirable a more complete connection between the physical mechanism behind dark energy and a fundamental theory of nature, one must bear in mind that an eternally accelerating universe, a rather generic feature of quintessence scenarios, seems not to be in agreement with String/M-theory (possible candidates of a model for quantum gravity) predictions, since it is endowed with a cosmological event horizon which prevents the construction of a
conventional S-matrix describing particle interactions \cite{fischler}. Although the transition from an initially decelerated to a late-time accelerating expansion is becoming
observationally established \cite{ms}, the duration of the accelerating phase, depends crucially on the cosmological scenario and, several models, which includes our current standard $\Lambda$CDM scenario, imply an eternal acceleration or even an accelerating expansion until the onset of a cosmic singularity (e.g., the so-called phantom cosmologies \cite{ph}). This dark energy/String theory conflict, therefore, leaves us with the formidable task of either finding alternatives to the conventional S-matrix or constructing a quintessence model of the Universe that predicts the possibility of a transient acceleration phenomenon. In this regard, an interesting quintessence scenario whose accelerating phase is a transient phenomenon has been proposed in Ref. \cite{prl}. In what follows, we highlight some of its features.

Let us first consider a homogeneous, isotropic, spatially flat cosmologies described by the FRW flat line element. The action for the model is given by 
\begin{equation}
\label{action}
S=\frac{m^2_{pl}}{16\pi}\int d^4 x \sqrt{-g}[R - {1\over2}\partial^{\mu}\Phi\partial_{\mu}\Phi-V(\Phi)+{\cal{L}}_{m}]\; ,
\end{equation}
where $R$ is the Ricci scalar and $m_{pl}\equiv G^{-1/2}$ is the Planck mass. The scalar field is assumed to be homogeneous, such that $\Phi=\Phi(t)$ and the Lagrangian density ${\cal{L}}_{m}$ includes all matter and radiation fields.

By combining the following \emph{ansatz} on the scale factor derivative of the energy density
\begin{equation}
\label{ansatz}
\frac{1}{\rho_{\Phi}}\frac{\partial\rho_{\Phi}}{\partial a}=-\frac{\lambda}{a^{1-2\alpha}}
\end{equation}
with the conservation equation for the quintessence component, i.e., $\dot\rho_{\Phi}+3H(\rho_{\Phi}+p_{\Phi})=0$, the expressions for the scalar field and its potential can be written as
\begin{equation}
 \label{phi1} 
\Phi(a) - \Phi_0 = \frac{1}{\sqrt{\sigma}}\ln_{1-\alpha}(a)\;, 
\end{equation}
and 
\begin{equation}
\label{gpotential} 
V(\Phi)= f(\alpha; \Phi) \rho_{\Phi,0}\exp\left[-\lambda\sqrt{\sigma}\left(\Phi + {\alpha \sqrt{\sigma} \over 2} \Phi^2 \right)  \right]\;.
\end{equation}
In the above expressions $\alpha$ and $\lambda $ are positive parameters, $\Phi_0$ is the current value of the field $\Phi$, $\sigma = {8\pi/\lambda m_{\rm pl}^2}$, $f(\alpha; \Phi) = [1-{\lambda\over6}(1+\alpha\sqrt{\sigma}\Phi)^2]$, and the generalized function $\ln_{1 - \xi}$, defined as $\ln_{1 - \xi}(x)\equiv{(x^{\xi}-1)/\xi}$, reduces to the ordinary logarithmic function in the limit $\xi \rightarrow 0$ \cite{abramowitz}. The important aspect to be emphasized at this point is that in the limit $\alpha \rightarrow 0$ Eqs. (\ref{phi1}) and (\ref{gpotential}) fully reproduce the exponential potential studied by Ratra and Peebles in Ref. \cite{RatraPeebles}, while $\forall$ $\alpha \neq 0$ the scenario described above represents a generalized model which admits a wider range of solutions.

The EoS for this quintessence component, i.e., 
\begin{equation}
\label{eq_state} 
w(a)= -1 + {\lambda \over 3}a^{2\alpha}\;, 
\end{equation}
is shown as a function of the redshift parameter ($z = a^{-1} - 1$) in Fig. (2a) for some selected values of the index $\alpha$ and $\lambda = 0.1$. The EoS above (which must lie in the interval $-1 \leq w(a) \leq 1$) is an increasingly function of time, being $\simeq -1$ in the past, $\simeq -0.96$ today, and becoming more positive in the future ($0$ at $a = 30^{1/2\alpha}$ and $1/3$ at $a = 40^{1/2\alpha}$). Such a behavior is typical of the so-called \emph{thawing} fields, as discussed in Refs. \cite{linder}. For this kind of fields, Ref. \cite{linder} also provides the following constraint
\begin{equation}
\label{thaws_out} 
1+ w < w'< 3(1 + w)\;, 
\end{equation}
where $w'= dw/d\ln{a}$. Note that, if it is natural to impose such a constraint at the epoch when dark energy starts becoming important, i.e., at $z \simeq 1$, then the above interval for $w'$ can be translated into the following bounds on the index $\alpha$
\begin{equation}
\label{alphathaws_out} 
{1}/{2} < \alpha < {3}/{2}\;. 
\end{equation}
Clearly, for all the values of $\alpha$ ranging in the above interval, the cosmic acceleration is a transient phenomenon since the quintessence field $\Phi$ will behave more and more as an attractive matter field. In order to better visualize this transient behavior, we show in Fig. (2b) the deceleration parameter, $q = -a\ddot{a}/\dot{a}^2$,  as a function of the redshift for some values of the index $\alpha$ and $\Omega_{m,0} = 0.27$. As can be seen from this figure, $\forall$ $\alpha \neq 0$ the Universe was decelerated in the past, began to accelerate at $z_a \lesssim 1$, is currently accelerated but will eventually decelerate in the future. Note also that, while the acceleration redshift $z_a$ depends very weakly on the value of $\alpha$, the deceleration redshift $z_d$ is strongly dependent, with the latter transition becoming more and more delayed as $\alpha \rightarrow 0$. As mentioned earlier, a cosmological behavior like the one described above seems to be in agreement with the requirements of String/M-theory (as discussed in Refs. \cite{fischler}), in that the current accelerating phase is a transitory phenomenon\footnote{Another interesting example of transient acceleration is provided by the brane-world scenarios discussed in Ref. \cite{ss} and the generalized EoS of Ref. \cite{stefancic}).}. As one may also check, the cosmological event horizon, i.e., the integral $\int{da/a^2H(a)}$ diverges for this transient scalar-field-dominated universe, thereby allowing the construction of a conventional S-matrix describing particle interactions within the String/M-theory frameworks. A typical example of an eternally accelerating universe, i.e., the $\Lambda$CDM model, is also shown in Fig. (2b) for the sake of comparison.

\section{Extra Dimensions}

So far we have discussed the phenomenon of cosmic acceleration as the result of unknown physical processes involving new fields in high energy physics. However, another possible route to deal with this dark \emph{pressure} problem could be a modification in gravity instead of any adjustment to the energy content of the Universe. This idea naturally brings to light another important question at the interface of fundamental physics and cosmology: extra dimensions. As is well known the existence of extra dimensions is required in various theories beyond the standard model of particle physics, especially in theories for unifying gravity and the other fundamental forces, such as String or M theories\footnote{Examples of modified gravity models also include scenarios with higher order curvature invariant modifications of the Einstein-Hilbert action, in which the natural matter dilution in the expanding universe is avoided by adding high order terms to the gravitational sector of the theory \cite{carroll1}.}. Extra dimensions may also provide a possible explanation for the so-called hierarchy problem, i.e.,  the huge difference between the electroweak and Planck scales [$m_{pl}/m_{EW} \sim 10^{16}$] \cite{hierarchy}. In this regard, an interesting scenario was proposed by Randall and Sundrum who showed that, if our 3-dimensional world is embedded in a 4-dimensional anti-de-Sitter bulk, gravitational excitations are confined close to our sub-manifold, giving rise to the familiar $1/r^2$ law of gravity \cite{rs}.

In the cosmological context, the role of extra spatial dimensions is translated into the so-called brane world (BW) cosmologies \cite{brane}. Many attempts to observationally detect or distinguish brane effects from the usual dark energy physics have been recently discussed in the literature. In Ref. \cite{ss}, for instance, Sahni and Shtanov investigated a class of BW models which admit a wider range of possibilities for the dark pressure than do the usual dark energy scenarios, while Maia et al., in Ref. \cite{maia}, showed that the dynamics of a dark energy component parametrized by a constant EoS ($w$) can be fully described by the effect of the extrinsic curvature of a FRW universe embedded into a 5-dimensional, constant curvature de-Sitter bulk.

\begin{figure*}[t]
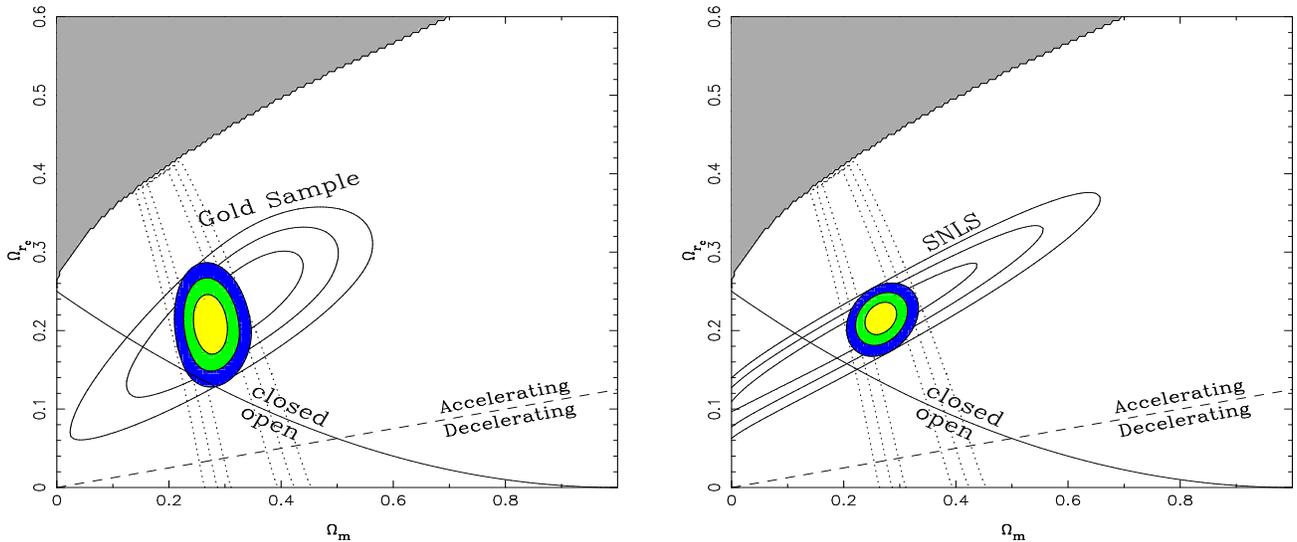

\centerline{\psfig{figure=f2.eps,width=3.2truein,height=2.8truein}
\hskip 0.3in
\psfig{figure=f3.eps,width=3.2truein,height=2.8truein}
\hskip 0.1in} 
\caption{{\bf{Left:}}  Probability contours at 68.3\%, 95.4\% and 99.7\% confidence levels for $\Omega_m$ versus $\Omega_{r_c}$ in the DGP model from the gold sample of SN Ia data (solid contours) and in conjunction with the SDSS baryon acoustic oscillations (coloured contours). The upper-left shaded region represents the ``no-big-bang" region, the thick solid line represents the flat universe and accelerated models of the universe are above the the dashed line. The best fit happens at $\Omega_m=0.272$ and $\Omega_{r_c}=0.211$. {\bf{Right:}} The same as in the previous Panel for an analysis involving  the first year SNLS data (solid contours) and in conjunction with the SDSS baryon acoustic oscillations (coloured contours). The best fit happens at $\Omega_m=0.265$ and $\Omega_{r_c}=0.216$ \emph{(Figures taken from \cite{zhuacl})}.}
\end{figure*}

Another interesting BW scenario is the one proposed by Dvali et al. \cite{dgp}, widely refered to as DGP model  (see also \cite{lue05} for a recent  review on the DGP
phenomenology). It describes a self-accelerating 5-dimensional BW model with a noncompact, infinite-volume extra dimension whose dynamics of gravity is governed by a competition between a 4-dimensional Ricci scalar term, induced on the
brane, and an ordinary 5-dimensional Einstein-Hilbert action, i.e., 
\begin{equation}
S = \frac{M_{(5)}^3}{2}\int{d^5y\sqrt{|{{g}}_{(5)}|}R_{(5)}} + \frac{m_{pl}^2}{2}\int{d^4x\sqrt{|g|}R}\; ,
\end{equation}
where $M_{(5)}$ denotes the 5-dimensional reduced Planck mass. The Friedmann equation which comes from this gravitational action coupled to matter on the brane take the form \cite{deff1}
\begin{equation} 
\left[\sqrt{\frac{\rho}{3M_{pl}^{2}} + \frac{1}{4r_{c}^{2}}} +
\frac{1}{2r_{c}}\right]^{2} = H^{2}  
+ \frac{k}{R(t)^{2}},
\end{equation} 
where $\rho$ is the energy density of the cosmic fluid, $k = 0, \pm 1$ is the spatial curvature and $r_c = M_{pl}^{2}/2M_{5}^{3}$ is the crossover scale defining the gravitational interaction among particles located on the brane. For scales below the crossover radius $r_c$ (where the induced 4-dimensional Ricci scalar dominates), the gravitational force experienced by two punctual sources is the usual 4-dimensional $1/r^{2}$ force, whereas for distance scales larger than $r_c$ the gravitational force follows the 5-dimensional $1/r^{3}$ behavior\footnote{As is well known, according to Gauss' law the gravitational force falls off as $r^{1 - S}$ where $S$ is the number of spatial dimensions.}. From the above equation it is also possible to define the density parameter associated with the crossover radius $r_c$, i.e., 
\begin{equation}
\Omega_{\rm{r_c}} = 1/4r_c^{2}H_o^{2}\; .
\end{equation} 
Note that estimates of $\Omega_{\rm{r_c}}$ means estimates on the length scale $r_c$. As shown by several authors \cite{dgpobs}, for values of $r_c \simeq H^{-1}_o$, the presence of an infinite-volume extra dimension as described above leads to a late-time acceleration of the Universe, in agreement with
most of the current distance-based cosmological observations\footnote{As noticed in Ref. \cite{deff1}, the above described cosmology can be exactly reproduced by the standard one plus an additional dark energy component with a time-dependent  equation of state parameter $\omega^{eff}(z) = 1/{\cal{G}}(z, \Omega_{\rm{m}},\Omega_{\rm{r_c}}) - 1$, where ${\cal{G}}(z, \Omega_{\rm{m}},\Omega_{\rm{r_c}}) = \sqrt{4\Omega_{\rm{r_c}}/\Omega_{\rm{m}}x'^{-3} + 4})(\sqrt{\Omega_{\rm{r_c}}/\Omega_{\rm{m}}x'^{-3}} + \sqrt{\Omega_{\rm{r_c}}/\Omega_{\rm{m}}x'^{-3} + 1})$ and $x' = (1 + z)^{-1}$.}. 

In Figure 3a we show the joint confidence contours at 68.3\%, 95.4\%, and 99.7\% confidence levels in the parametric space  $\Omega_{\rm{m}}- \Omega_{\rm{r_c}}$ arising from the \emph{gold} Sne Ia sample \cite{rnew} and the SDSS baryon acoustic oscillations \cite{bao} (see \cite{zhuacl,ariel} for more details). The best-fit parameters for this analysis are $\Omega_{\rm{m}} = 0. 272$ and $\Omega_{\rm{r_c}} = 0.211$. Note that the best-fit value for $\Omega_{\rm{r_c}}$  leads to an estimate of the crossover scale $r_c$ in terms of the Hubble radius $H_o^{-1}$, i.e., $r_c \simeq 1.089 H_o^{-1}$ . Figure 3b illustrates the allowed regions in the $\Omega_{\rm{m}}- \Omega_{\rm{r_c}}$ plane by using the first-year SNLS data \cite{Legacy2005} in conjunction with the SDSS baryon acoustic oscillations. The best fit for this joint SNLS plus BAO analysis happens at  $\Omega_{\rm{m}} =0.265$ and  $\Omega_{\rm{r_c}} = 0.216$, which is very closed to the WMAP estimates for the clustered matter\footnote{It is also worth emphasizing that in most of the observational analyses of DGP model a spatially closed universe is largely favoured, which seems to be in disagreement with current WMAP results indicating a nearly flat universe with $\Omega_k \simeq -0.02$ \cite{wmap}.}. From these and other recent results (see \cite{dgpobs} for more observational analyses in DGP models) we note that this class of BW models provide a good description for the current obserbational data, which may be thought of as an indication that the existence of extra dimensions play an important role not only in fundamental physics but also in cosmology.

\section{Conclusions}

The results of observational cosmology in the past decade have opened up an unprecedented opportunity to establish a more solid connection between fundamental physics and cosmology. Surely, the most remarkable finding among these results comes from SNe Ia observations which suggest that the cosmic expansion is undergoing a late time acceleration. These SNe Ia results have been checked and confirmed by other cosmological observables, including the angular distribution of the 3-K CMB temperature and the clustering and dynamical estimates on the matter density parameter, which certainly makes a serious case for dark energy. Theoretical efforts to explain the nature of this dark energy component are vast and include, besides the possibilities discussed here, models with mass varying neutrinos \cite{neutrinos}, holographic dark energy \cite{elcio}, two-field quintom models \cite{quintom}, models driven by dissipative process \cite{mc}, unification dark matter/dark energy \cite{cg}, among many others. 

Here, we have explored three specific scenarios of dark energy and shown that, although completely different from the physical viewpoint, all of them have attractive features capable of explaining the current cosmic acceleration and other recent astronomical observations. We emphasize that a possible way to distinguish among some of these dark energy candidates is through the observational limits on the cosmic EoS ($w$) and its time derivative ($w'$), i.e., if upcoming observations confirrm $w = 1$, then one will need to review the $\Lambda$ problem, in order to fathom out why the formally infinite quantity $\rho_{\Lambda}$ is in fact so very small. On the other hand, if either $w \neq −1$ or it is a time-dependent quantity($w' \neq 0$) , then one will be able to rule out the cosmological constant as the actual nature of the dark energy and search for more realistic quintessence,  BW and other dark energy models. We expect that future observations along with theoretical developments will be able to decide this matter and shed some light on the origin and nature of the dark energy.

\vspace{0.2cm}

\begin{acknowledgments}

I wish to thank the organizers of the XXVI ENFPC for the opportunity to attend such an interesting meeting. I am also very grateful to J. A. S. Lima, N. Pires, R. Silva, Z.-H. Zhu, D. Jain, A. Dev and F. C. Carvalho for valuable discussions. This work is supported by by CNPq under Grants No. 307860/2004-3 and 475835/2004-2, and by Funda\c{c}\~ao de Amparo \`a Pesquisa do Estado do Rio de Janeiro (FAPERJ), No. E-26/171.251/2004.
\end{acknowledgments}


\begin{thebibliography}{30}



\bibitem{perl}S. perlmutter et al. Nature {\bf 391}, 51 (1998).

\bibitem{rnew} A. G. Riess et al., Astrophys. J. {\bf{607}} 665 (2004)

\bibitem{Legacy2005} P. Astier et al., Astron. Astrophys. {\bf{447}}, 31 (2006). 

\bibitem{age} J. S. Dunlop et al., Nature {\bf{381}}, 581 (1996); B. Chaboyer et al., Astrophys. J. {\bf{494}}, 96 (1998); R. Cayrel et al., Nature {\bf{409}}, 691 (2001); G. Hasinger et al., Astrophys. J., 573, L77 (2002); A. Fria\c{c}a , J. S. Alcaniz  and J. A. S. Lima, Mon. Not. Roy. Astron. Soc. {\bf{362}}, 1295 (2005). astro-ph/0504031.

\bibitem{boom} P. de Bernardis {et al.}, Nature {\bf{404}}, 955 (200)

\bibitem{wmap}  D. N. Spergel et al., \apj Suppl. {\bf{148}}, 175 (2003); D. N. Spergel et al., 2006. \texttt{astro-ph/0603449}.

\bibitem{calb} R. G. Calberg {\it et al.}, Astrophys. J. {\bf{462}}, 32 (1996)


\bibitem{revde} V. Sahni and A. A. Starobinsky, Int. J. Mod. Phys. D9, 373 (2000); P. J. E. Peebles and B. Ratra Rev. Mod. Phys. {\bf{75}}, 559 (2003); T. Padmanabhan, Phys. Rept. {\bf{380}}, 235 (2003); J. A. S. Lima, BJP {\bf 34}, 194 (2004); D. H. Weinberg, New Astron. Rev. {\bf{49}}, 337 (2005). astro-ph/0510196; E. J. Copeland et al, hep-th/0603057.

\bibitem{zeldo}  Zel'dovich, Ya. B., Zh. Eksp. Teor. Fiz., Pis'ma Red. {\bf{6}}, 883 (1967) [JETP Lett. 6, 316 (1967)]; Zel’dovich, Ya.B., Sov. Phys. – Uspekhi {\bf{11}}, 381 (1968).


\bibitem{lambda} S. Weinberg, Rev. Mod. Phys. 61, 1 (1989); S. M. Carroll, Living Rev. Relativity 4,  (2001).

\bibitem{paddy} T. Padmanabhan, in `` Graduate School in Astronomy: X Special Courses at the National Observatory",  Eds. S. Daflon, J.S. Alcaniz, E. Telles, R. de la Reza, AIP Conf .Proc. {\bf{843}}, p. 111 (2006).

\bibitem{ozer} M. $\ddot{\rm{O}}$zer and M. O. Taha, Phys. Lett. B 171, 363 (1986); Nucl. Phys. B287, 776 (1987);  O. Bertolami, Nuovo Cimento Soc. Ital. Fis., B 93, 36 (1986).

\bibitem{bron} M. Bronstein, Phys. Z. Sowjetunion {\bf{3}} (1933).

\bibitem{lambdat} K. Freese et al., Nucl. Phys. {\bf{B287}}, 797 (1987); W. Chen and Y-S. Wu, Phys. Rev. {\bf D41}, 695 (1990);  J. C. Carvalho, J. A. S. Lima and I. Waga, Phys. Rev. {\bf{D46}} 2404 (1992); I. Waga, Astrophys. J. {\bf{414}}, 436 (1993); J. A. S. Lima and J. M. F. Maia, Phys. Rev {\bf D49}, 5597 (1994); J. A. S. Lima and M. Trodden, Phys. Rev. {\bf D53}, 4280 (1996); A. I. Arbab and A. M. M. Abdel-Rahman, Phys.Rev. {\bf D50}, 7725 (1994); L .F. Bloomfield Torres and I. Waga, Mon. Not. Roy. Astron. Soc. {\bf{279}}, 712 (1996); O. Bertolami and P. J. Martins, Phys. Rev. {\bf{D61}}, 064007 (2000); R. G. Vishwakarma, GRG {\bf{33}}, 1973 (2001) A. S. Al-Rawaf, Mod. Phys. Lett. {\bf A14}, 633 (2001);  J. V. Cunha, J. A. S. Lima and J. S. Alcaniz, Phys. Rev. {\bf{D66}}, 023520 (2002). astro-ph/0202260; J. S. Alcaniz and J. M. F. Maia, Phys. Rev. {\bf{D67}}, 043502 (2003). astro-ph/0212510; S. Carneiro and J. A. S. Lima, IJMP {\bf A20} 2465 (2005);  S. Carneiro et al. Phys. Rev. {\bf{D74}}, 023532 (2006).  astro-ph/0605607.

\bibitem{overduin} J. M. Overduin and F. I. Cooperstock, Phys. Rev. {\bf{D58}}, 043506 (1998).



\bibitem{wm} P. Wang and X. Meng, Class. Quant. Grav. {\bf 22}, 283 (2005).

\bibitem{alcaniz} J. S. Alcaniz and J. A. S. Lima, Phys. Rev. {\bf{D72}}, 063516 (2005). astro-ph/0507372

\bibitem{shapiro}  I. L. Shapiro and J. Sola, Phys. Lett. {\bf{B475}}, 236 (2000); I. L. Shapiro and J. Sola, JHEP, {\bf 0202}, 006 (2002);  J. Sola and H. Stefancic, astro-ph/0507110; I. L. Shapiro, J. Sola, C. Espana-Bonet and P. Ruiz-Lapuente, Phys. Lett. {\bf{B574}}, 149 (2003); I. L. Shapiro, J. Sola and H. Stefancic, JCAP {\bf{0501}}, 012 (2005).

\bibitem{ms} M. S. Turner and A. G. Riess, Astrophys. J. {\bf{569}}, 18 (2002).

\bibitem{sarkar} M. Birkel and S. Sarkar, Astropart. Phys. {\bf{6}}, 197, (1997).


\bibitem{vamp} J. A. Casas, J. Garcia-Bellido and M. Quir\'os, Class. Quant. Grav. {\bf 9}, 1371 (1992); G. W. Anderson, S. M. Carroll, astro-ph/9711288; L. Amendola, MNRAS {\bf 342}, 221 (2003); M. Pietroni, Phys. Rev. {\bf D67}, 103523 (2003); G. R. Farrar and P. J. E. Peebles, \apj {\bf{604}}, 1 (2004).

\bibitem{allen} S. W. allen et al. MNRAS {\bf{360}}, 546, (2005).

\bibitem{hst} W. L. Freedman {et al.}, Astrop. J. {\bf{553}}, 47 (2001).


\bibitem{refer}T. Padmanabhan and T. R. Choudhury, Mon. Not. Roy. Astron. Soc. {\bf{344}}, 823 (2003); J. A. S. Lima, J. V. Cunha and J. S. Alcaniz, Phys. Rev. {\bf{D68}}, 023510 (2003). astro-ph/0303388;  P. T. Silva and O. Bertolami, \apj {\bf{599}}, 829 (2003); Z.-H. Zhu and M.-K. Fujimoto, \apj {\bf{585}}, 52 (2003); S. Nesseris and L. Perivolaropoulos, Phys. Rev. {\bf{D70}}, 043531 (2004); Y. Wang and M. Tegmark, Phys. Rev. Lett. {\bf{92}}, 241302 (2004); J. V. Cunha, J. A. S. Lima and J. S. Alcaniz, Phys. Rev. {\bf{D69}}, 083501 (2004).astro-ph/0306319; J. S. Alcaniz and N. Pires, Phys. Rev. {\bf{D70}}, 047303 (2004). astro-ph/0404146; T. R. Choudhury and T. Padmanabhan, Astron. Astrophys. {\bf{429}} 807 (2005);  H. K. Jassal, J. S. Bagla and T. Padmanabhan, Phys. Rev. {\bf{D72}}, 103503 (2005); D. Rapetti, S. W. Allen and J. Weller, Mon. Not. Roy. Astron. Soc. {\bf{360}}, 546 (2005); R. Lazkoz, S. Nesseris and L. Perivolaropoulos, JCAP {\bf{0511}}, 010 (2005); B. Feng, X. Wang and X.  Zhang, Phys. Lett. {\bf{B607}}, 35 (2005); A. Shafieloo, U. Alam, V. Sahni and A. A. Starobinsky,  astro-ph/0505329; M. A. Dantas et al.,  astro-ph/0607060. H. K.Jassal, J. S. Bagla, T. Padmanabhan, astro-ph/0601389.

\bibitem{quintessence}  R. R. Caldwell, R. Dave, P. J. Steinhardt, Phys. Rev. Lett. {\bf 80}, 1582 (1998); S. M. Carroll, Phys. Rev. Lett., {\bf{81}}, 3067 (1998);  L.-M. Wang, R. R. Caldwell, J. P. Ostriker, and P. J. Steinhardt, Astrophys. J. {\bf 530}, 17 (2000); J. S. Bagla , H. K. Jassal and T. Padmanabhan, Phys. Rev. {\bf{D67}}, 063504 (2003); K. R. S. Balaji and R. H. Brandenberger, Phys. Rev. Lett. {\bf{94}}, 031301 (2005);  A. Albrecht and C. Skordis, Phys. Rev. Lett. {\bf{84}}, 2076 (2000); L. R. W. Abramo and F. Finelli, Phys. Lett. {\bf{B575}}, 165 (2003); C. Rubano et al., Phys. Rev. D{\bf 69}, 103510 (2004); V.~Faraoni and M.~N.~Jensen, Class.\ Quant.\ Grav.\  {\bf 23}, 3005 (2006).

\bibitem{RatraPeebles}  B. Ratra and P. J. E. Peebles, Phys. Rev D{\bf 37}, 3406 (1988).

\bibitem{fj} P.G. Ferreira and M. Joyce, Phys. Rev. D{\bf 58}, 023503(1998).

\bibitem{wetterich} C. Wetterich, Astron. \& Astrophys. {\bf{301}}, 321 (1995).

\bibitem{power_law} P.J.E. Peebles and B. Ratra, Astrophys. J. Lett. {\bf 325}, L17 (1988); I. Zlatev, L-M. Wang , P.J. Steinhardt, Phys. Rev. Lett. {\bf{82}}, 896 (1999).

\bibitem{dodelson} S. Dodelson et al., M. Kaplinghat and E. Stewart,  Phys. Rev. Lett. {\bf{85}}, 5276 (2000).

\bibitem{fischler} W. Fischler, A. Kashani-Poor, R. McNees, and S. Paban, JHEP {\bf{3}}, 0107 (2001); S. Hellerman, N. Kaloper and L. Susskind, JHEP {\bf{3}}, 0106,
(2001); J.M. Cline, JHEP 0108, 35 (2001);  E. Halyo, JHEP 0110, 025 (2001).

\bibitem{ph} R. R. Caldwell, Phys. Lett. B \textbf{545}, 23 (2002);  V. Faraoni, Int. J. Mod. Phys. {\bf{D11}}, 471 (2002);   S. M. Carroll, M. Hoffman and M. Trodden,  Phys. Rev. {\bf{D68}},  023509 (2003); R. R. Caldwell, M. Kamionkowski and N. N. Weinberg, Phys. Rev. Lett. {\bf{91}}, 071301 (2003);  P. F. Gonzalez-Diaz, Phys. Rev. {\bf{D68}}, 021303 (2003); J. S. Alcaniz, Phys. Rev. {\bf{D69}}, 083521 (2004). astro-ph/0312424; S. Nesseris and L. Perivolaropoulos, Phy. Rev. {\bf{D70}}, 123529 (2004); J. A. S. Lima and J. S. Alcaniz, Phys. Lett. {\bf{B600}}, 191 (2004).  astro-ph/0402265;  F. C. Carvalho and A. Saa, Phys. Rev. {\bf{D70}}, 087302 (2004); J. Santos  and J. S. Alcaniz, Phys. Lett. {\bf{B619}}, 11 (2005). astro-ph/0502031; L. R. l. Abramo and N. Pinto-Neto, Phys. Rev. {\bf{D73}}, 063522 (2006); E. M. Barboza Jr. and N. A. Lemos, gr-qc/0606084.

\bibitem{prl} F. C. Carvalho, J. S. Alcaniz, J. A. S. Lima and R. Silva, Phys. Rev. Lett. {\bf{97}}, 081301 (2006). astro-ph/0608439.


\bibitem{abramowitz} M. Abramowitz and I. Stegun, {\em Handbook of Mathematical Functions}, Dover, New York, (1965). See also J.A.S. Lima, R. Silva and A.R. Plastino, Phys. Rev. Lett., {\bf 86}, 2938 (2001).


\bibitem{linder} R. R. Caldwell and E. V. Linder, Phys. Rev. Lett. {\bf{95}}, 141301 (2005); R. J. Scherrer, Phys. Rev. {\bf{D73}}, 043502 (2006); T. Chiba, Phys. Rev. {\bf{D73}}, 063501 (2006); V. Barger, E. Guarnaccia and D. Marfatia, Phys. Lett. {\bf{B635}}, 61 (2006).

\bibitem{ss} V. Sahni and Y. Shtanov, IJMP {\bf{D11}}, 1515 (2002);  V. Sahni and Y. Shtanov, JCAP {\bf{0311}}, 014 (2003).

\bibitem{stefancic} J. S.  Alcaniz and H. Stefancic, astro-ph/0512622.

\bibitem{carroll1} S. M. Carroll, V. Duvvuri, M. Trodden and M. S. Turner, Phys.Rev. D 70, 043528 (2004); S. M. Carroll et al., Phys. Rev. {\bf{D71}}, 063513 (2005).

\bibitem{hierarchy} N. Arkani-Hamed, S. Dimopoulos, G. R. Dvali, Phys. Lett. {\bf{B429}}, 263 (1998); I. Antoniadis, N. Arkani-Hamed, S. Dimopoulos and G. Dvali, Phys. Lett. {\bf{B436}} 257 (1998).

\bibitem{rs} L. Randall and R. Sundrum, Phys. Rev. Lett. {\bf{83}}, 3370 (1999); Phys. Rev. Lett. {\bf{83}}, 4690 (1999)

\bibitem{brane} T. Shiromizu, K. Maeda, and M. Sasaki, Phys. Rev. D 62, 024012 (2000); R. Dick, Class. Quantum Grav. 18, R1 (2001);  C. J. Hogan, Class. Quant. Grav. {\bf{18}}, 4039 (2001); C. J. Hogan, astro-ph/0104105; K. Ichiki, M. Yahiro, T. Kajino, M. Orito and G. J. Mathews, Phys. Rev. {\bf{D66}}, 043521 (2002); K. Freese and M. Lewis, Phys. Lett. {\bf{B540}}, 1 (2002); Z. -H. Zhu and M. -K. Fujimoto, \apj, {\bf{581}}, 1 (2002); D. Langlois, Astrophys. Space Sci. {\bf{283}}, 469 (2003); Z. -H. Zhu and M. -K. Fujimoto, \apj {\bf{585}}, 52 (2003); Z. -H. Zhu and M. -K. Fujimoto, \apj {\bf{602}}, 12 (2004). H. Zhang and Z.-H. Zhu, astro-ph/0607531 



\bibitem{maia} M. D. Maia, E. M. Monte and J. M. F. Maia, Phys. Lett. {\bf{B585}}, 11 (2004). astro-ph/0208223; M. D. Maia, E. M. Monte, J. M. F. Maia and J. S. Alcaniz, Class. Quant. Grav. {\bf{22}}, 1623 (2005). astro-ph/0403072.

\bibitem{dgp} G. Dvali, G. Gabadadze and M. Porrati, Phys. Lett. {\bf{B485}}, 208 (2000).


\bibitem{lue05} A. Lue, Phys. Rept. {\bf{423}}, 1 (2006).

\bibitem{deff1} C. Deffayet et al., Phys. Rev. {\bf{D66}}, 024019 (2002)

\bibitem{dgpobs} C. Deffayet, G.  Dvali and G. Gabadadze, Phys. Rev. {\bf{D65}}, 044023 (2002); P. P. Avelino and C. J. A. P. Martins, ApJ {\bf{565}}, 661 (2002); J. S. Alcaniz, Phys. Rev. D {\bf{65}}, 123514 (2002). astro-ph/0202492;  D. Jain, A. Dev and J. S. Alcaniz, Phys. Rev. {\bf{D66}}, 083511 (2002). astro-ph/0206224; J. S. Alcaniz, D. Jain and A. Dev, Phys. Rev. {\bf{D66}}, 067301 (2002). astro-ph/0206448; A. Lue, Phys. Rev. {\bf{D67}}, 064004 (2003);  A. Lue and G. D. Starkman, Phys. Rev. {\bf{D67}}, 064002 (2003);  A. Lue, R. Scoccimarro and G. D. Starkman,  Phys. Rev. {\bf{D69}}, 124015; E. V. Linder, Phys. Rev. Lett. {\bf{90}}, 091301 (2003); Z. -H. Zhu and J. S. Alcaniz, Astrophys. J. {\bf{620}}, 7 (2005). astro-ph/0404201; J. S. Alcaniz and Z. -H. Zhu, Phys. Rev. {\bf{D71}}, 083513 (2005). astro-ph/0411604; N. Pires, Z.-H. Zhu and J. S. Alcaniz, Phys. Rev. {\bf{D73}}, 123530 (2006). astro-ph/0606689.


\bibitem{bao} D. J. Eisenstein et al., Astrophys.J. {\bf{633}}, 560 (2005).

\bibitem{zhuacl} Z.-K. Guo, Z.-H. Zhu, J. S. Alcaniz and Y.-Z. Zhang, Astrophys.J. {\bf{646}}, 1 (2006).

\bibitem{ariel} M. Fairbairn and A.l Goobar, astro-ph/0511029.


\bibitem{neutrinos}  P. Gu, X. Wang and X. Zhang, Phys. Rev. {\bf{D68}}, 087301 (2003); R. Fardon, A. E. Nelson, N. Weiner, JCAP {\bf{0410}}, 005 (2004); D. B. Kaplan, A. E. Nelson and N. Weiner, Phys. Rev. Lett. {\bf{93}}, 091801 (2004).

\bibitem{elcio} M. Li, Phys. Lett. {\bf{B603}}, 1 (2004); S. D. H. Hsu, Phys. Lett. {\bf{B594}}, 13 (2004); B. Wang, Y.-G. Gong and E. Abdalla, Phys. Lett. {\bf{B624}}, 141 (2005); B. Wang, C.-Y. Lin and E. Abdalla, Phys. Lett. {\bf{B637}}, 357 (2006); Z.-Y. Huang, B. Wang, E. Abdalla and R.-K. Su, JCAP {\bf{0605}}, 013 (2006); J. P. B. Almeida and J. G. Pereira, Phys. Lett. {\bf{B636}}, 75 (2006); B. Wang, J. Zang, C.-Y. Lin, E. Abdalla and S. Micheletti, astro-ph/0607126; W. Zimdahl and D. Pavon, astro-ph/0606555. 


\bibitem{quintom} B. Feng, X.L. Wang and X. Zhang, Phys. Lett. {\bf{B607}}, 35 (2005);  Z. K. Guo, Y. S. Piao, X. Zhang and Y. Z. Zhang, Phys. Lett. {\bf{B608}}, 17 (2005).

\bibitem{mc} A. Kamenshchik, U. Moschella and V. Pasquier, Phys. Lett. {\bf{B511}}, 265 (2001); N. Bilic, G. B. Tupper and R. D. Violler, Phys. lett. {\bf{B535}} 17 (2002); M. C. Bento, O Bertolami and A. A. Sen, Phys. Rev. {\bf{D66}}, 043507 (2002); A. Dev, J. S. Alcaniz and D. Jain, Phys. Rev {\bf{D67}}, 023515 (2003). astro-ph/0209379; J. S. Alcaniz, D. Jain and A. Dev, Phys. Rev. {\bf{D67}}, 043514 (2003). astro-ph/0210476; M. Makler, S. Q. de Oliveira and I. Waga, Phys. Lett. {\bf{B555}}, 1 (2003); D. Carturan and F. Finelli, Phys. Rev. {\bf{D68}}, 103501(2003);  R. R. R. Reis, I. Waga,  M. O. Calvao and S. E. Joras, Phys. Rev. {\bf{D68}}, 061302 (2003); Z.-H. Zhu, Astron. Astrophys. {\bf{423}}, 421 (2004); O. Bertolami, A. A. Sen, S. Sen and P. T. Silva, MNRAS {\bf{353}}, 329 (2004);  A. Dev,  D. Jain and J. S. Alcaniz, Astron. Astrophys. {\bf{417}}, 847 (2004). astro-ph/0311056; M.C. Bento, O. Bertolami, M.J. Rebouças and P.T. Silva, Phys. Rev. {\bf{D73}}, 043504 (2006); H. Zhang and Z.-H. Zhu, Phys. Rev. {\bf{D73}}, 043518 (2006); J. A. S. Lima, J. V. Cunha and J. S. Alcaniz, astro-ph/0608469.

\bibitem{cg} I. Prigogine, J. Geheniau, E. Gunzig and P. Nardone, Gen. Relat. Gravit. {\bf{21}}, 767 (1989); M. O. Calvao, J. A. S. Lima and I. Waga, Phys. Lett. {\bf{A162}}, 223 (1992);  J. A. S. Lima and A. S. M. Germano, Phys. Lett. {\bf{A170}}, 373 (1992); J. Triginer, W. Zimdahl and D. Pavon, Class. Quant. Grav. {\bf{13}}, 403 (1996); J. A. S. Lima, J. S. Alcaniz, Astron. Astrophys. {\bf{348}}, 1 (1999). astro-ph/9902337; J. S. Alcaniz, J. A. S. Lima, Astron. Astrophys. {\bf{349}}, 729 (1999).astro-ph/9906410;  J. A. S. Lima, A. S. M. Germano, L. R. W. Abramo, Phys. Rev. {\bf{D53}}, 4287 (1996); L. P. Chimento, A. S. Jakubi,  D. Pavon and W. Zimdahl, Phys. Rev. {\bf{D67}}, 083513 (2003). Y. Qiang and T-J. Zhang, astro-ph/0503123.




\end{thebibliography}

\end{document}